\documentclass[intlimits,twoside,a4paper]{article}

\usepackage{amsmath,amssymb,amsthm}
\usepackage{graphicx}

\usepackage[T2A]{fontenc}
\usepackage[cp1251]{inputenc}

\usepackage{cmpj2}

\issue{2013}{16}{2}{23603}

\doinumber{10.5488/CMP.16.23603}

\title[Universality versus nonuniversality in asymmetric fluid criticality]%
{Universality versus nonuniversality in asymmetric fluid criticality}

\author[M.A. Anisimov]{M.A. Anisimov}
 \address{Institute for Physical Science and Technology, University of Maryland,
College Park, MD 20742, USA}

\date{Received  January 14, 2013, in final form March 27, 2013}
\authorcopyright{M.A. Anisimov, 2013}

\begin{document}

\maketitle

\begin{abstract}
Critical phenomena in real fluids demonstrate a combination of universal
features caused by the divergence of long-range fluctuations of density and
nonuniversal (system-dependent) features associated with specific
intermolecular interactions. Asymptotically, all fluids belong to the
Ising-model class of universality. The asymptotic power laws for the
thermodynamic properties are described by two independent universal critical
exponents and by two independent nonuniversal critical amplitudes; other
critical amplitudes can be obtained by universal relations. The nonuniversal
critical parameters (critical temperature, pressure, and density) can be
absorbed in the property units. Nonasymptotic critical behavior of fluids
can be divided into two parts, symmetric (``Ising-like'') and asymmetric
(``fluid-like''). The symmetric nonasymptotic behavior contains a new
universal exponent (Wegner exponent) and the system-dependent crossover
scale (Ginzburg number) associated with the range of intermolecular
interactions, while the asymmetric features are generally described by an
additional universal exponent and by three nonasymptotic amplitudes
associated with mixing of the physical fields into the scaling fields.
\end{abstract}

\maketitle

\section{Introduction}

Universality of critical phenomena is one of the most fascinating concepts
in physics of condensed matter \cite{fisher1982,domb1996} \ Phase
transitions of strikingly different nature, such as para-ferro-magnetism,
vaporization, or fluid demixing may be described by the same equation of
state near the critical points if a proper (``isomorphic'') set of thermodynamic
variables is chosen. There are several classes of universality defined
through the dimension of the order parameter. The order parameter may be a
scalar, $\mathit{n}$-component vector or a tensor. For example, the order
parameter in fluids is associated with the density or concentration (a
scalar) while the order parameter in anisotropic magnetics (magnetization)
is a one-component vector. The Ising model of anisotropic ferromagnets is
mathematically equivalent to the lattice-gas model which describes
the condensation of fluids. In the both cases, the order parameter is
one-dimensional ($n=1$). The isomorphism between the members of a universality
class can be established by mapping the thermodynamic variables of one
system onto another. In addition, the order parameter can be conserved (such
as density) or non-conserved (such as magnetization). This particular nature
of the order parameter affects the phase-transition dynamics.

It is well established, primarily through experiments \cite{Anisimov
book,sengers2009}, \ that all fluids and fluid mixtures belong to the
Ising-model class of universality in statics and to the
conserved-order-parameter class of universality in dynamics. This
universality is associated with the universal nature of critical
fluctuations. The fluctuations of the order parameter diverge at the
critical point. The correlation length of the order-parameter fluctuations
becomes much larger than  the range of intermolecular interactions, thus
making the details of the intermolecular potential unimportant. Landau and
Lifshitz stated in an earlier edition of ``Statistical
Physics'' \cite{LL 1958 edition}, \textit{``Unlike solids and gases, liquids do not allow a general calculation of their thermodynamics quantities or even their temperature dependence. The reason
for this is the presence of strong interactions between the molecules of the
liquid without having, at the same time, the smallness of vibrations which
makes the thermal motions of solids so simple.''}
Undoubtedly, this statement is not applicable to liquids in the vicinity of
their critical points. Thermodynamic properties in the critical region of
very different substances, such as helium isotopes and other inert gases,
organic liquids and water can be theoretically predicted; they are all
described by the universal power laws, also known as scaling laws, which are
characterized by the universal critical exponents.

However, there are still non-universal features in the critical behavior of
fluids. The critical parameters (temperature, pressure, and density) are
obviously non-universal, being determined by specific intermolecular
potentials. The critical temperature ranges from a few kelvins for helium
isotopes to thousands of kelvins for liquid metals. This particular
non-universal feature can be eliminated by reducing the properties of a
particular substance by its critical parameters. Another non-universal
feature is the size of the asymptotic critical region in which the universal
scaling laws are valid. This size is controlled by the so-called Ginzburg
number which depends on the range of intermolecular interactions. Finally,
real fluids, unlike the lattice-gas/Ising model, are asymmetric with respect
to the critical isochor. The fluid asymmetry causes additional specific
non-asymptotic corrections to the universal critical behavior. In this
paper, I present a brief overview of universal and nonuniversal
contributions to the equation of sate of near-critical fluids.

\section{Universal asymptotic criticality}

T{\normalsize he fluctuation-induced non-analytic critical behavior can be
asymptotically described by scaling theory in terms of two independent
scaling fields, namely, }${h}_{{1}}${  \
(``ordering'' field) and }${h}_{2}${  \ (``thermal''
field) and two conjugate scaling densities, namely, the order parameter }$%
\phi _{1}$ (strongly fluctuating){  \ and }$\phi _{2}$ (weakly
fluctuating){  . The third field, }$h_{3}$, is the critical part
of an appropriate field-dependent thermodynamic potential, which is defined
as a function exhibiting a minimum at equilibrium with respect to a
variation of the order parameter. The differential of the third field is
\begin{equation}
\rd h_{3}=\phi _{1}\rd h_{1}+\phi _{2}\rd h_{2}\,.  \label{dh3}
\end{equation}
In the scaling theory, the field potential $h_{3}$ is a homogeneous function of
${h}_{{1}}$ and ${h}_{2}$. Asymptotically,
\begin{equation}
h_{3}\approx \left\vert h_{2}\right\vert ^{2-\alpha }f^{\pm }\left( \frac{%
h_{1}}{\left\vert h_{2}\right\vert ^{2-\alpha -\beta }}\right) ,
\label{h3_2}
\end{equation}%
where $f^{\pm }$ is a scaling function and the superscript $\pm $ refers to $%
h_{2}>0$ and $h_{2}<0$, respectively. Here and below, $\approx $ means
asymptotically equal, while $\simeq $ means approximately equal. The
critical point is defined by the condition $h_{1}=h_{2}=h_{3}=0$. The form
of the scaling function is universal; however, it contains two
thermodynamically independent (but system-specific) amplitudes. All other
asymptotic amplitudes are related to the selected ones by universal
relations. The critical exponents $\alpha $ and $\beta $ are universal
within a class of critical-point universality. All fluids and fluid mixtures
belong to the Ising-model universality class. The Ising values for $\alpha
\simeq 0.109$ and $\beta \simeq 0.326$, are well established theoretically
and confirmed experimentally \cite{Anisimov
book,Sengers:ARPC1986, Liu:PA1989, Guida98, Campostrini:PRE1999,Campostrini:PRE1999_,
Pelissetto:PR2002,FisherZinn98,Straub:1999,Anisim:book2000,sengers2009,Chapter 14,Chapter 10}. Two Ising amplitudes, $\hat{A}_{0}$ and $\hat{B}_{0}$ are determined by
the asymptotic power-law behavior of the two scaling densities in zero
ordering field ($h_{1}=0$):
\begin{eqnarray}
\phi _{1} &=&\left( \frac{\partial h_{3}}{\partial h_{1}}\right)
_{h_{2}}\approx \pm \hat{B}_{0}\left\vert h_{2}\right\vert ^{\beta }\text{ }%
\left( h_{2}<0\right) ,  \label{phi1} \\
\phi _{2} &=&\left( \frac{\partial h_{3}}{\partial h_{2}}\right)
_{h_{1}}\approx \frac{\hat{A}_{0}^{\pm }}{1-\alpha }h_{2}\left\vert
h_{2}\right\vert ^{-\alpha }+B_{\mathrm{cr}}|h_{2}|,  \label{phi2}
\end{eqnarray}%
and of the three scaling susceptibilities, ``strong'' $\chi _{1}$,  ``weak'' $\chi _{2}$, and ``cross'' $\chi _{12}$ in zero
ordering field:
\begin{eqnarray}
\chi _{1} &=&\left( \frac{\partial \phi _{1}}{\partial h_{1}}\right)
_{h_{2}}\approx \hat{\Gamma}_{0}^{\pm }\left\vert h_{2}\right\vert ^{-\gamma
},  \label{chi1} \\
\chi _{2} &=&\left( \frac{\partial \phi _{2}}{\partial h_{2}}\right)
_{h_{1}}\approx \hat{A}_{0}^{\pm }\left\vert h_{2}\right\vert ^{-\alpha },
\label{ch2} \\
\chi _{12} &=&\left( \frac{\partial \phi _{1}}{\partial h_{2}}\right)
_{h_{1}}\approx \beta \hat{B}_{0}\frac{\left\vert h_{2}\right\vert ^{\beta }%
}{h_{2}}\qquad\left( h_{2}<0\right) ,  \label{chi3}
\end{eqnarray}%
where the strong susceptibility critical exponent
\begin{equation}
\gamma =2-\alpha -2\beta \simeq 1.239,
\end{equation}%
and the strong susceptibility critical amplitude $\hat{\Gamma}_{0}^{\pm }$
is related to $\hat{B}_{0}$ and $\hat{A}_{0}^{\pm }$ through universal
ratios as \cite{FisherZinn98}:
\begin{eqnarray}
\alpha \hat{\Gamma}_{0}^{+}\hat{A}_{0}^{+}/\hat{B}_{0}^{2} &\simeq &0.0581,
\\
\hat{\Gamma}_{0}^{+}/\hat{\Gamma}_{0}^{-} &\simeq &4.8, \\
\hat{A}_{0}^{+}/\hat{A}_{0}^{-} &\simeq &0.523.
\end{eqnarray}%
While the superscript $\pm $ refers to the states at $h_{2}>0$ and $%
h_{2}<0, $ the prefactor $\pm $ in equation~(\ref{phi1}) refers to the two
branches of the order parameter corresponding to $h_{1}\,>0$ and $h_{1}\,<0$
sides (in the limit $h_{1}=0)$ , respectively. The field-dependent potential
$h_{3}$ is symmetric with respect to the sign of the ordering field $h_{1}$
and, hence, to the sign of the order parameter $\phi _{1}$. In these
expressions $\hat{A}_{0}^{\pm }$, $\hat{\Gamma}_{0}^{\pm }$, and $\hat{B}%
_{0} $ are non-universal critical amplitudes. The term in $\phi _{2}$
proportional to $B_{\mathrm{cr}} $ is an analytic fluctuation-induced
contribution to the second scaling density \cite{Anisimov1992}. Strictly
speaking, this term is not asymptotic since the term $h_{2}\left\vert
h_{2}\right\vert ^{-\alpha }$ dominates.

Additional universal relations connect the critical exponent $\nu \simeq
0.63 $ of the correlation length (diverging in zero ordering field as $\xi
=\xi _{0}^{\pm }\left\vert h_{2}\right\vert ^{-\nu })$ and $\alpha $,
\begin{equation}
2-\alpha =d\nu
\end{equation}%
(where $d$ is the number of space dimensions), and the amplitudes $\hat{A}%
_{0}^{+}$ and $\xi _{0}^{+}$,
\begin{equation}
A_{0}^{+}\rho _{\mathrm{c}}\left( \xi _{0}^{+}\right) ^{3}\simeq 0.172.
\end{equation}%
This relation is known as the two-scale factor of universality \cite%
{fisher1982,FisherZinn98}. The ratio $\xi _{0}^{+}/\xi _{0}^{-}\simeq 1.96$
is also universal \cite{FisherZinn98}.

In the mean-field approximation, with $\alpha =0$ and $\beta =1/2$, equation~(\ref%
{h3_2}) reduces to the asymptotic Landau expansion \cite{Lifshitz:book1980},%
\begin{equation}
-h_{3}\approx \frac{1}{2}a_{0}h_{2}\phi _{1}^{2}+\frac{1}{24}u_{0}\phi
_{1}^{4}-h_{1}\phi _{1}\,,  \label{lggrandpotential}
\end{equation}%
where $a_{0}$ and $u_{0}$ are mean-field system-dependent amplitudes. The
amplitude $a_{0}$ is  unimportant. It can be eliminated by rescaling the
fields $h_{3}$ and $h_{1}$ and the coupling constant $u_{0}\rightarrow
u=u_{0}/a_{0}$.

In the lattice-gas model, the ordering field $h_{1}$ is associated with the
chemical potential $\mu ,$ the thermal field $h_{2}$ is associated with the
temperature $T$, and the order parameter is associated with the molecular
density $\rho ,$ while $h_{3}$ is associated with the pressure $P$. In both,
the scaling regime and the mean-field approximation, the thermodynamic
properties of the lattice gas are symmetric with respect to the sign of the
order parameter. Similar to the lattice-gas model, in real one-component
fluids, the thermodynamic fields are the temperature $T$, the chemical
potential $\mu $, and the pressure $P$, while the conjugate densities are
the number density $\rho $ and the entropy density $s=\rho S$ ($S$ is the
entropy per molecule). The physical variables are interrelated by the
Gibbs-Duhem relation
\begin{equation}
\rd P=\rho \rd\mu +s\rd T.  \label{gd P}
\end{equation}%
Consequently, the densities, namely, the molecular density and the entropy
density are derived from the pressure as
\begin{equation}
\rho =\left( \frac{\partial P}{\partial \mu }\right) _{T}\,,\qquad s=\left(
\frac{\partial P}{\partial T}\right) _{\mu }\,.
\end{equation}%
In addition to the reduced density $\Delta \hat{\rho}$ and reduced
temperature $\Delta \hat{T}$,
\begin{equation}\label{eq:17}
\Delta \hat{\rho}=\frac{\rho -\rho _{\mathrm{c}}}{\rho _{\mathrm{c}}}\,
,\qquad \Delta \hat{T}=\frac{T-T_{\mathrm{c}}}{T_{\mathrm{c}}}\,,
\end{equation}%
it is convenient to define
\begin{equation}\label{eq:18}
\Delta \hat{s}=\frac{s-s_{\mathrm{c}}}{\rho _{\mathrm{c}}k_{\mathrm{B}}}\,%
,\qquad\Delta \hat{P}=\frac{P-P_{\mathrm{c}}}{\rho _{\mathrm{c}}k_{\mathrm{B}%
}T_{\mathrm{c}}}\,,\qquad\Delta \hat{\mu}=\frac{\mu -\mu _{\mathrm{c}}}{k_{%
\mathrm{B}}T_{\mathrm{c}}}\,,
\end{equation}%
where $k_{\mathrm{B}}$ is Boltzmann's constant. In equations~(\ref{eq:17})--(\ref{eq:18}) and below,
the subscript ``c'' denotes the properties at the critical point.

As shown by Anisimov et al. \cite{anisimov1995}, since
in classical thermodynamics, the absolute value of entropy is arbitrary, the
critical value of entropy can be chosen upon practical convenience. It is
 clearly seen from the basic thermodynamic relation
\begin{equation}
\rd P=\rho \rd\mu +\rho S\rd T,
\end{equation}%
that
\begin{equation}
\frac{\rd\hat{\mu}}{\rd\hat{T}}+\hat{S}_{\mathrm{c}}-\left( \frac{\partial \hat{P%
}}{\partial \hat{T}}\right) _{\mathrm{c}}=0.\
\end{equation}%
Thus, with adopting $\hat{S}_{\mathrm{c}}=( \partial \hat{P}/\partial
\hat{T}) _{h_{1}=0,\mathrm{c}}$, we obtain $( \partial \hat{\mu}%
/\partial \hat{T})_{h_{1}=0,\mathrm{c}}=0$ meaning that in the linear
approximation, the chemical potential along the vapor-liquid coexistence does
not depend on temperature. With this choice of the critical entropy, we
find for the critical part of pressure (the density of the grand
thermodynamic potential $\Omega =-PV$) after subtracting its regular part
\begin{equation}\label{eq:21}
\Delta \widetilde{P}=\Delta \hat{P}-\Delta \hat{\mu}-\hat{s}_{\mathrm{c}%
}\Delta \hat{T}.
\end{equation}%
Then, asymptotically, for one-component fluids, the scaling fields have the
following simple relations to the physical fields:%
\begin{align}
h_{1}& =\Delta \hat{\mu}, \\
h_{2}& =\Delta \hat{T}, \\
h_{3}& =\Delta \widetilde{P}.
\end{align}

One can conclude that in the asymptotic regime, in addition to the
system-dependent critical parameters, which can be eliminated by rescaling
the units of thermodynamic properties, there are only two independent
critical amplitudes. The independent critical amplitudes correspond to the
relevant nonuniversal coefficients, $u$ and $g$, in the asymptotic
Landau-Ginzburg Hamiltonian, given in terms of the spatially dependent order
parameter $\phi _{1}(\mathbf{x})$ as \cite{Lifshitz:book1980}
\begin{equation}
\mathcal{H}=\frac{1}{2}(\Delta \hat{T})\phi _{1}^{2}+\frac{u}{4!}\phi
_{1}^{4}+\frac{1}{2}g\phi _{1}\left( {\nabla} \phi _{1}\right) ^{2}-h_{1}\phi
_{1}\,.
\end{equation}

The universal scaling function $f^{\pm }$ in equation~(\ref{h3_2}) for practical
applications is commonly calculated from a parametric equation of state,
such as the linear model \cite{schofield1969,Chapter 10}, which has been
shown to be accurate to an order of $\epsilon ^{2}$ in the $\epsilon $-expansion,
where $\epsilon =4-d$ \cite{brezin1972}.

\section{Size of the critical region and symmetric corrections to asymptotic
scaling laws}

The universal scaling laws discussed in the previous Section are valid only
asymptotically, very close to the critical point. Upon departure from the
critical point, corrections to the asymptotic power laws appear. The first
correction, also known as the Wegner correction \cite{wegner1972} contains a
new universal scaling function, $g_{1}^{\pm }(z)$, and a new universal
critical exponent $\theta $\ (known as the ``Wegner exponent''). In the
first-order $\epsilon $-expansion $\theta =\epsilon /2$ \cite{wegner1972}.

The Wegner correction arises from the difference between the
renormalization-group fixed-point coupling constant $u^{\ast }$ and the
system dependent mean-field value of the coupling constant $u$. When the
Wegner correction is included, the Ising field-dependent potential [equation~(\ref{h3_2})] reads \cite{pelissetto2002}
\begin{equation}
h_{3}=|h_{2}|^{2-\alpha }f^{\pm }(z)1+|h_{2}|^{\theta }f_{1}^{\pm }(z)+\ldots,
\label{Wegner h3}
\end{equation}%
where $z=h_{1}/|h_{2}|^{2-\alpha -\beta }$ and $f_{1}^{\pm }\propto (u^{\ast
}-u)$. The nonasymptotic (``confluent'') scaling function $f_{1}^{\pm }$
contains additional system-dependent parameters, $(u^{\ast }-u)$ and the
Ginzburg number, $N_{\mathrm{G}}=u^{2}\big[ \rho _{\mathrm{c}}\left( \xi
_{0}^{+}\right) ^{3}\big] ^{-2}$.

Thus, the scaling power laws are to be complemented by confluent
singularities. In terms of physical variables in zero ordering field

\begin{equation}
\Delta \widetilde{P}=|\Delta \hat{T}|^{2-\alpha }\left[ 1+\left( \frac{%
|\Delta \hat{T}|}{t_{\times }}\right) ^{\theta }+\left( \frac{|\Delta \hat{T}%
|}{t_{\times }}\right) ^{2\theta }+\ldots\right] ,
\end{equation}%
where the crossover scale $t_{\times }\propto \left( N_{\mathrm{G}}\right)
(u^{\ast }-u)^{-2}$, which can be considered as the effective Ginzburg
number, defines the size of the asymptotic critical region, $|\Delta \hat{T}| \ll t_{\times }$. The amplitude $B_{\mathrm{cr}}$ in the analytic
fluctuation-induced contribution to $\phi _{2}$ given by equation~(\ref{phi2})
also depends on the crossover scale $t_{\times }$ \cite{Anisimov1992,
KimAnisimov2003}.

\section{Nonasymptotic asymmetry corrections}

The canonical mapping of the liquid-vapor critical point onto Ising
criticality is given by the lattice-gas model \cite{lee1952}. This model can
easily be extended to binary fluids and fluid mixtures through a
reassignment of variables and the principle of isomorphism \cite%
{anisimov1995}. For the remainder of the text, the liquid-vapor
one-component system is only discussed. The lattice-gas model preserves the
exact symmetry of uniaxial Ising-type ferromagnets and consequently, the
liquid-vapor coexistence curve of the lattice gas is symmetric with respect
to the density $\rho $. The order parameter of the lattice gas is the
reduced density, $\Delta \hat{\rho}=(\rho -\rho _{\mathrm{c}})/\rho _{%
\mathrm{c}}$. If the liquid and vapor branches of the coexistence curve are
denoted by \textquotedblleft $+$\textquotedblright\ and \textquotedblleft $-$%
\textquotedblright\ respectively, the asymmetric portion of the density is
given by the excess density
\begin{equation}
\Delta \hat{\rho}_{\mathrm{d}}=\frac{\Delta \hat{\rho}^{+}+\Delta \hat{\rho}%
^{-}}{2}\,.
\end{equation}%
For the lattice gas, $\Delta \hat{\rho}_{\mathrm{d}}=0$. However, real fluids
do not possess the symmetry of the Ising model, and in general $\Delta \hat{%
\rho}_{\mathrm{d}}\neq 0$. Even the coexistence curve of $^{3}$He, the most
symmetric fluid known, exhibits some small asymmetry \cite{hahn2004}. In
asymmetric systems, the leading behavior is still determined by the
Ising-type behavior, and asymmetric corrections appear as sub-leading terms
in the quantities like density. In mean-field models of the liquid-vapor
critical point, such as the van der Waals model, the asymmetry of the
coexistence curve is described by the ``law'' of rectilinear diameter~\cite{cailletet1886,cailletet1886_}
\begin{equation}
\Delta \hat{\rho}_{\mathrm{d}}=D_{1}|\Delta \hat{T}|,  \label{lin diam}
\end{equation}%
where the reduced temperature is defined by $\Delta \hat{T}=(T-T_{\mathrm{c}%
})/T_{\mathrm{c}}$, with $T_{\mathrm{c}}$ being the critical temperature.
While some one-component fluids such as xenon \cite{narger1990} seem to
asymptotically follow this ``law'', others,
like SF$_{6}$ \cite{weiner1974}, show strong deviations from rectilinearity
in the critical region.

Models such as the Widom-Rowlinson penetrable-sphere model \cite{widom1970}
and Mermin's decorated-lattice models \cite{mermin1971, mermin1971b} predict
non-classical, {i.e.}, non-mean-field, behavior of the excess
density. On the basis of these models, a non-classical theory of fluid
criticality, known as ``revised scaling''
\cite{rehr1973} was proposed. The formulation of a revised scaling postulates
that the Ising scaling fields are analytic functions of the chemical
potential $\mu $ and temperature $T$, whereas the lattice gas model assumes
that $\mu $ and $T$ are the correct scaling fields. This field mixing
produces the following asymptotic behavior:
\begin{equation}
\Delta \hat{\rho}_{\mathrm{d}}\approx D_{1-\alpha }|\Delta \hat{T}%
|^{1-\alpha }+D_{1}|\Delta \hat{T}|.  \label{rs diam}
\end{equation}

Additional theoretical support for a revised scaling came from Nicoll and Zia
\cite{nicoll1981}, and Nicoll \cite{nicoll1981b}, who performed a
field-theoretic (FT) analysis of an asymmetric Landau-Ginzburg-Wilson (LGW)
Hamiltonian and found that a revised scaling arises naturally from the
inclusion of asymmetric operators in the Hamiltonian. In addition, they
found that these asymmetric operators also lead to a non-analytic correction
to the excess density characterized by a new asymmetric
correction-to-scaling exponent $\theta _{5}$. The excess density predicted
by their analysis goes as
\begin{equation}
\Delta \hat{\rho}_{\mathrm{d}}\approx D_{1-\alpha }|\Delta \hat{T}%
|^{1-\alpha }+D_{1}|\Delta \hat{T}|+D_{\beta +\theta _{5}}|\Delta \hat{T}%
|^{\beta +\theta _{5}}.  \label{rg diam}
\end{equation}
The universal exponent $\theta _{5}$ was found to be $\theta
_{5}=1/2+\epsilon =3/2$ in the first-order $\epsilon $-expansion, where $%
\epsilon =4-d$ and $d$ is the spatial dimensionality \cite%
{leyKoo1981,vause1980,vause1980_}. Working to order $\epsilon ^{3}$, Zhang and Zia \cite%
{zhang1982}, found their results to be consistent with the bound $\theta
_{5}\gtrsim 1.0$.

More recently, Fisher and co-workers \cite{fisher2000,kim2003} have argued
for an extended formulation of scaling, originally discussed by Rehr and
Mermin \cite{rehr1973}, which is now known as ``complete scaling''. This
theory of asymmetric fluid criticality is an extension of the field-mixing
in a revised scaling and incorporates the hypothesis of Griffiths and Wheeler
\cite{griffiths1970} that preferable thermodynamic variables do not exist.
This concept implies that pressure $P$, chemical potential $\mu$, and
temperature $T$ should all be treated on equal footing in any formulation of
scaling for the liquid-vapor critical point. The Ising scaling fields should,
therefore, be treated as analytic functions of all three. By contrast,
a revised scaling assigns a special role to the pressure $P(\mu,T)$ as the
field-dependent thermodynamic potential. A complete scaling predicts that the
excess density is asymptotically given by
\begin{equation}
\Delta\hat{\rho}_{\mathrm{d}}\approx D_{2\beta}|\Delta\hat{T}%
|^{2\beta}+D_{1-\alpha}|\Delta\hat{T}|^{1-\alpha}+D_{1}|\Delta\hat{T}|,
\label{cs diam}
\end{equation}
where $2\beta\simeq0.65$. This result clearly differs from the FT
prediction, equation~(\ref{rg diam}). In the mean-field approximation, the
connection between a complete scaling and the asymmetric Landau expansion has
been investigated by Anisimov and Wang \cite{anisimov2006,wang2007}, who
demonstrated that the two approaches appear to be consistent. A complete
scaling has also been extended to inhomogeneous fluids by Bertrand and
Anisimov \cite{bertrand2010}. That the penetrable-sphere model does not
exhibit complete scaling, has been investigated by Ren {et al.} \cite{ren2006}, who found that this is due to a special symmetry of the model.

In addition to the leading $2\beta $ term in the excess density, a complete
scaling also predicts a divergence in the second derivative of the chemical
potential along the coexistence curve
\begin{equation}
\left( \frac{\rd^{2}\mu }{\rd T^{2}}\right)_{\mathrm{cxc}}\sim |\Delta \hat{T}%
|^{-\alpha },
\end{equation}%
where the subscript $\mathrm{cxc}$ denotes the conditions of phase
coexistence. The so-called Yang-Yang anomaly derives its name from the
Yang-Yang relation \cite{yang1964}
\begin{equation}
\frac{\rho C_{V}}{T}=\left( \frac{\rd^{2}P}{\rd T^{2}}\right) _{\mathrm{cxc}%
}-\rho \left( \frac{\rd^{2}\mu }{\rd T^{2}}\right) _{\mathrm{cxc}},
\label{yangyang}
\end{equation}%
where $C_{V}$ is the isochoric heat capacity. A complete scaling implies that the
divergence of the isochoric heat capacity is shared between the second
derivatives of the pressure and the chemical potential. By contrast, a revised
scaling predicts that $(\rd^{2}\mu /\rd T^{2})_{\mathrm{cxc}}$ remains finite at
the critical point. Nicoll's analysis also predicts a non-analytic behavior of
the chemical potential, specifically,
\begin{equation}
\left( \frac{\rd^{2}\mu }{\rd T^{2}}\right) _{\mathrm{cxc}}\sim |\Delta \hat{T}%
|^{-\alpha -\beta +\theta _{5}},
\end{equation}%
however, the relatively large value of $\theta _{5}$ ensures that this
quantity remains finite at the critical point.

Fisher and co-workers have found support for a complete scaling in heat
capacity measurements \cite{orkoulas2000} and computer simulations of highly
asymmetric fluid models \cite%
{orkoulas2001,kim2003b,kim2003c,kim2005,kim2005b}. Anisimov and Wang have
demonstrated that a complete scaling is also supported by the data on liquid-vapor
coexistence in highly asymmetric fluids \cite{anisimov2006,wang2007}. There
is also at least one model that exhibits the type of field mixing
characteristic of complete scaling \cite{felderhof1970,felderhof1970_}. A complete scaling
remains, however, an essentially phenomenological theory.

\section{Complete scaling}

\label{complete scaling}

For the liquid-vapor transition, the principle of complete scaling asserts
that the scaling fields can be expanded in $\Delta \hat{\mu}$, $\Delta \hat{T%
}$, and $\Delta \hat{P}$. In the lowest order approximation, the scaling
fields are given by
\begin{align}
h_{1}& \simeq a_{1}\Delta \hat{\mu}+a_{2}\Delta \hat{T}+a_{3}\Delta \hat{P},
\label{h1 tmp} \\
h_{2}& \simeq b_{1}\Delta \hat{T}+b_{2}\Delta \hat{\mu}+b_{3}\Delta \hat{P},
\label{h2 tmp} \\
h_{3}& \simeq c_{1}\Delta \hat{P}+c_{2}\Delta \hat{\mu}+c_{3}\Delta \hat{T}%
+c_{23}\Delta \hat{\mu}\Delta \hat{T},  \label{h3 tmp}
\end{align}%
where the constant coefficients are called mixing coefficients. In general,
the complete scaling transformations should include terms of all orders in $%
\Delta \hat{T}$, $\Delta \hat{\mu}$, and $\Delta \hat{P}$. Here, we only
consider contributions to the excess density $\Delta \hat{\rho}_{\mathrm{d}}$
which are of the order of $|\Delta \hat{T}|$ or lower. Two second-order terms
satisfy this criterion, $\Delta \hat{\mu}\Delta \hat{T}$ when added to $%
h_{3} $ and $\Delta \hat{T}^{2}$ when added to $h_{1}$ and $h_{3}$. However,
we have omitted explicit $\Delta \hat{T}^{2}$ terms from the relations for $%
h_{1}$ and $h_{3}$ since these can be absorbed into the regular, i.e., non-critical, portion of the thermodynamic potential without affecting our
results. The exact connection between the transformations, equations~(\ref{h1 tmp}%
)--(\ref{h3 tmp}), and the excess density will be derived in the following
paragraphs. Once this connection is established, one can verify that the
remaining second-order terms $\Delta \hat{\mu}\Delta \hat{P}$, $\Delta \hat{P%
}\Delta \hat{T}$, $\Delta \hat{P}^{2}$, and $\Delta \hat{\mu}^{2}$ do not
need to be included in this approximation.

As discussed by Wang and Anisimov \cite{wang2007} and Bertrand \cite%
{bertrand2011}, the transformations, equations~(\ref{h1 tmp})--(\ref{h3 tmp}), can
be much simplified by selecting normalizations for the scaling
fields, adopting a particular value of $\hat{s}_{\mathrm{c}}=s_{\mathrm{c}%
}/\rho _{\mathrm{c}}k_{\mathrm{B}}$, which is arbitrary in classical
thermodynamics, and neglecting higher order terms. Specifically, we choose $%
\hat{s}_{\mathrm{c}}=(\rd P/\rd T)_{h_{1}=0,\mathrm{c}}$. These simplifications can
be implemented by adopting the following choice of coefficients
\begin{align}
&a_{1}=(1-a),& &a_{2}=-a\hat{s}_{\mathrm{c}},& &a_{3}=a,& &
\label{simp_coeff1} \\
&b_{1}=1,& &b_{2}=b,& &b_{3}=0,& & \\
&c_{1}=1,& &c_{2}=-1,& &c_{3}=-\hat{s}_{\mathrm{c}},& &c_{23}=c.
\label{simp_coeff3}
\end{align}%
When these coefficients are substituted into the complete scaling
transformations, we find that the transformations reduce to
\begin{align}
h_{1}& =\Delta \hat{\mu}+a\Delta \widetilde{P},  \label{h1 simp} \\
h_{2}& =\Delta \hat{T}+b\Delta \hat{\mu},  \label{h2 simp} \\
h_{3}& =\Delta \widetilde{P}+c\Delta \hat{\mu}\Delta \hat{T},
\label{h3 simp}
\end{align}%
where $\Delta \widetilde{P}$ is defined by equation~(\ref{eq:21}). In the mean-field
approximation, $\Delta \hat{\mu}\sim |\Delta \hat{T}|^{3/2}$ and $\Delta
\hat{P}\sim |\Delta \hat{T}|^{2}$, so that each asymmetric term in the
complete scaling transformations is smaller than the leading term by a
factor of $|\Delta \hat{T}|^{1/2}$. The revised scaling transformations are
reproduced in the absence of pressure mixing ($a=0$), and the lattice gas
model is recovered when all mixing coefficients are set to zero ($a=b=c=0$).

The physical densities can be found in terms of the scaling densities from equations~(\ref{h1 simp})--(\ref{h3 simp}) with the result
\begin{align}
&\Delta \hat{\rho}=\frac{\phi _{1}+b\phi _{2}-c\Delta \hat{T}}{1-a\phi _{1}}\,,
\\
&\Delta \hat{s}=\frac{\phi _{2}}{1-a\phi _{1}}\,,
\end{align}%
where $\phi _{1}$ and $\phi _{2}$ are given by equations~(\ref{phi1}) and (\ref%
{phi2}), respectively. To the leading order in the asymmetry and reduced
temperature, these expressions are given by
\begin{align}
&\Delta \hat{\rho}\simeq \phi _{1}+a(\phi _{1})^{2}+b\phi _{2}-c\Delta \hat{T}%
,  \label{cs rho} \\
&\Delta \hat{s}\simeq \phi _{2}\,.
\end{align}%
When the scaling densities presented in equations~(\ref{phi1}) and (\ref{phi2})
are substituted into equation (\ref{cs rho}), the complete scaling excess density
introduced in equation (\ref{cs diam}) is reproduced with the coefficients
\begin{align}
D_{2\beta }&=a(B_{0})^{2}\,, \\
D_{1-\alpha }&=-b\frac{A_{0}^{-}}{1-\alpha }\,, \\
D_{1}&=B_{\mathrm{cr}}+c\,.
\end{align}%
We note that the leading $2\beta $ term is proportional to the pressure
mixing coefficient $a$. The same is true of the Yang-Yang anomaly, which
follows from the first complete scaling relationship, equation~(\ref{h1 simp}),
as
\begin{equation}
\left( \frac{\rd^{2}\hat{\mu}}{\rd\hat{T}^{2}}\right) _{\mathrm{cxc}}\simeq
-a\left( \frac{\rd^{2}\hat{P}}{\rd\hat{T}^{2}}\right) _{\mathrm{cxc}%
}=-aA_{0}^{-}|\Delta \hat{T}|^{-\alpha },  \label{mu cxc}
\end{equation}%
where, to the leading order, the coexistence curve is defined by $h_{1}=0$.

Complete scaling also predicts the effects of fluid asymmetry on other
thermodynamic properties. In particular, the physical susceptibilities, such
as the isothermal compressibility, volumetric expansivity, and the heat
capacity are found to be combinations of all three scaling susceptibilities:
\textquotedblleft strong\textquotedblright\ $\chi _{1}$, \textquotedblleft
weak\textquotedblright\ $\chi _{2}$, and \textquotedblleft
cross\textquotedblright\ $\chi _{12}$ \cite{wang2007}.

\section{Discussion and conclusion}

As shown in previous section, asymmetric fluid criticallity generally
introduces three additional, nonuniversal and independent, amplitudes
associated with the mixing of physical fields into the Ising scaling fields.
However, as the comparison between a complete scaling and the FT approach to
asymmetric fluid criticality shows \cite{chris2012}, the complete scaling
and FT equations of state are nearly identical, except that the FT equation
of state has an additional term responsible for the asymmetric
correction-to-scaling exponent $\theta _{5}$.

As a result, the asymtery-induced excess density can be written
\begin{eqnarray}
\Delta \hat{\rho}_{\mathrm{d}}\approx D_{1-\alpha }|\Delta \hat{T}%
|^{1-\alpha }+D_{2\beta }|\Delta \hat{T}|^{2\beta }+D_{1}|\Delta \hat{T}|+D_{\beta +\theta _{5}}|\Delta \hat{T}|^{\beta +\theta
_{5}},
\end{eqnarray}%
where $D_{\beta +\theta _{5}}\propto u_{5}^{\mathrm{eff}},$ a new
nonasymptotic amplitude. For many practical applications, the contribution
from $\theta _{5}$ can be neglected. In this regime, the complete scaling
and FT approaches are equivalent. In practice, the number of independent
amplitudes may be constrained by a particular equation of state.

There is an analogy between the asymmetric correction-to-scaling exponent $%
\theta _{5}$ and the Wegner correction-to-scaling exponent $\phi \simeq
\epsilon /2$ \cite{wegner1972}. The Wegner correction arises from the
difference between the renormalization-group fixed-point coupling constant $%
u^{\ast }$ and the system-dependent mean-field value of the coupling
constant $u$ [{cf.} equation~(\ref{Wegner h3})]. As in the case of the
Wegner correction, which is associated with an additional critical amplitude
$u^{\ast }-u$, the $\theta _{5}$ exponent is associated with the new
critical amplitude $u_{5}^{\mathrm{eff}}$ which is the difference between
the fifth-order coefficient in the asymmetric Landau expansion and the
amplitude of the asymmetry of the gradient term in the effective
Hamiltonian \cite{chris2012}. If $u_{5}^{\mathrm{eff}}=0$, $h_{3}$ includes
only the leading asymmetric terms. In this particular case, complete scaling
becomes exact. After the complete scaling has taken care of the leading
asymmetric corrections by the mixing of physical fields into the scaling
field, the field dependent potential could be extended as
\begin{equation}
h_{3}\approx |h_{2}|^{2-\alpha }f^{\pm }(z)\left[ 1+|h_{2}|^{\theta
}f_{1}^{\pm }(z)+|h_{2}|^{\theta _{5}}f_{\mathrm{asym}}^{\pm }(z)\right] ,
\end{equation}%
where $f_{\mathrm{asym}}^{\pm }\propto u_{5}^{\mathrm{eff}}$. However, there
is a significant difference between these two corrections-to-scaling. Unlike
$\theta _{5}$, the exponent $\theta $ vanishes in the mean-field
approximation $\epsilon =0$. This explains why the Wegner correction can be
consistently omitted in the mean-field approximation. The same is not true
of $\theta _{5}$, because in the mean-field approximation $\theta _{5}=1/2$.

\section*{Acknowledgements}
I thank C.E.~Bertrand, J.F.~Nicoll, and J.V.~Sengers for collaboration
and M.E. Fisher for discussions and comments. I also appreciate long-term
fruitful interactions with scientists from the Institute for Condensed
Matter Physics, the National Academy of Sciences of Ukraine, in particular
with I.R.~Yukhnovskii and M.P.~Kozlovskii who made important contributions
to the physics of critical phenomena and phase transitions~\cite{Yukhnovskii,Kozlovskii}.

\ukrainianpart

\title%
{Універсальність чи неуніверсальність   в асиметричній критичності плинів}
\author{М.А. Анісімов}
 \address{Інститут фізичної науки і технології, Університет Мариленду, Коледж Парк, MD 20742, США}

\makeukrtitle

\begin{abstract}
\tolerance=3000%
Критичні явища в реальних плинах демонструють комбінацію
універсальних рис, спричинених роз\-біжністю далекосяжних флуктуацій
густини, і неуніверсальних (системо залежних) рис, пов'язаних із
специфічними міжмолекулярними взаємодіями. Асимптотично всі плини
належать до класу універсальності моделі Ізинга. Асимптотичні
степеневі закони для термодинамічних властивостей описуються двома
незалежними універсальними критичними показниками і двома
незалежними неуніверсальними критичними амплітудами; решту критичних
амплітуд можна отримати з універсальних співвідношень.
Не\-універсальні критичні параметри  (критична температура, тиск і
густина) можуть бути включені в одиниці цих властивостей.
Неасимптотичну критичну поведінку плинів можна поділити на дві
частини, симетричну (``ізингоподібну'') і асиметричну
(``плиноподібну''). Симетрична не\-асимптотична поведінка містить новий
універсальний показник (показник Вегнера) і системо залеж\-ний масштаб
кросоверу (число Гінз\-бурга), пов'язаний з областю дії
міжмолекулярних взаємодій, тоді як асиметричні риси взагальному
описуються додатковим універсальним показником і трьома
неасимптотичними амплітудами, пов'язаними зі змішуванням фізичних
полів у скейлінгових полях.

\keywords плини, критична точка, універсальність, повний скейлінг
\end{abstract}

\end{document}